\documentclass[aps,prd,superscriptaddress,amsmath,twocolumn,10pt]{revtex4-1}
\usepackage{color,hangcaption,hhline,psfrag,rotating,amssymb}
\usepackage[hang,nooneline]{subfigure}
\usepackage{dcolumn}
\usepackage{verbatim}
\usepackage{bm}

\begin{document}
\title{The size of the pion from full lattice QCD with physical $u$, $d$, $s$ and 
$c$ quarks}

\author{J.~Koponen}
\email[]{jonna.koponen@glasgow.ac.uk}
\affiliation{SUPA, School of Physics and Astronomy, University of Glasgow, Glasgow, G12 8QQ, UK}
\author{F.~Bursa}
\affiliation{SUPA, School of Physics and Astronomy, University of Glasgow, Glasgow, G12 8QQ, UK}
\author{C.~T.~H.~Davies}
\email[]{christine.davies@glasgow.ac.uk}
\affiliation{SUPA, School of Physics and Astronomy, University of Glasgow, Glasgow, G12 8QQ, UK}
\author{R.~J.~Dowdall}
\affiliation{DAMTP, University of Cambridge, Wilberforce Road, Cambridge, CB3 0WA, UK}
\author{G.~P.~Lepage}
\affiliation{Laboratory of Elementary-Particle Physics, Cornell University, Ithaca, New York 14853, USA}

\collaboration{HPQCD collaboration}
\homepage{http://www.physics.gla.ac.uk/HPQCD}
\noaffiliation

\date{\today}

\begin{abstract}

We present the first calculation of the electromagnetic form 
factor of the $\pi$ meson at physical light quark masses. 
We use configurations generated by the MILC collaboration 
including the effect of $u$, $d$, $s$ and $c$ sea quarks 
with the Highly Improved Staggered Quark formalism. 
We work at three values of the lattice spacing on large volumes and with 
$u$/$d$ quark masses going down to the physical value. 
We study scalar and vector form factors for a range in 
space-like $q^2$ from 0.0 to -0.1 $\mathrm{GeV}^2$ 
and from their shape 
we extract mean square radii. Our vector form factor 
agrees well with experiment and 
we find $\langle r^2 \rangle_V = 0.403(18)(6) \,\mathrm{fm}^2$. 
For the scalar form factor we include quark-line 
disconnected contributions which have a significant 
impact on the radius. We give the first results for 
SU(3) flavour-singlet and 
octet scalar mean square radii, obtaining: 
$\langle r^2 \rangle_S^{\mathrm{singlet}} = 0.506(38)(53) \mathrm{fm}^2$ and  
$\langle r^2 \rangle_S^{\mathrm{octet}} = 0.431(38)(46) \mathrm{fm}^2$. 
We discuss the comparison with expectations from chiral perturbation 
theory. 

\end{abstract}

% insert suggested PACS numbers in braces on next line
%\pacs{}
% insert suggested keywords - APS authors don't need to do this
%\keywords{}

%\maketitle must follow title, authors, abstract, \pacs, and \keywords
\maketitle

\section{Introduction} 
\label{sec:intro}

The electromagnetic form factor of the charged $\pi$ meson 
parameterises the deviations from the behaviour of a point-like 
particle when struck by a photon. These deviations result from 
the internal structure of the $\pi$ i.e. its quark constituents 
and their strong interaction. 
The form factor is calculable in QCD but a fully nonperturbative 
treatment is necessary at the small (negative) values of 4-momentum transfer, $q^2$, 
covered by direct model-independent experimental determination of the 
vector form factor~\cite{amendolia} 
from $\pi-e$ scattering. 
The experimental error is 1-1.5\% in the region up to $|q^2| = 0.1 \mathrm{GeV}^2$ 
and so a lattice QCD calculation of the form factor there can provide 
a stringent test of QCD. This is complementary to tests of QCD through 
calculation of meson masses and of decay constants 
that parameterise meson annihilation, for 
example to a $W$ boson~\cite{latqcd, fkpi}.    

In the nonrelativistic limit, where $q^2 \approx -(\vec{q})^2$,
the vector form factor, $f_+(q^2)$, can be viewed as the Fourier 
transform of the electric charge 
distribution. The mean squared radius obtained by 
integrating over this distribution is then 
given by 
\begin{equation}
\label{eq:rdef}
\langle r^2 \rangle  = 6 \frac{df_+(q^2)}{dq^2} \bigg|_{q^2=0}.
\end{equation}
This is adopted more generally as a definition of $\langle r^2 \rangle$,
since it is useful to have a single number 
with which to characterise the shape of the form factor. 
We will use it here to compare the `size' of 
the $\pi$ derived from our form factor with that obtained 
from experiment. 

The calculation of $\langle r^2 \rangle$ from lattice 
QCD is complicated by the fact that the result is very sensitive to 
the mass of the $\pi$. The mean square radius 
diverges as $m_{\pi}^2 \rightarrow 0$ 
when the $\pi$ meson cloud surrounding the $\pi$ becomes of 
infinite range~\cite{gasserl}. It has been numerically too 
expensive until recently to include $u/d$ quarks with their 
physically very light masses in lattice QCD calculations. 
Results have instead had to be extrapolated to the physical 
point from heavier masses using chiral perturbation theory. 
The lightest $\pi$ meson mass used in earlier calculations 
of the electromagnetic form factor has been in the range of 
250-400 MeV, i.e. approximately twice the physical value or more. 
Results range from multiple values of the lattice spacing including 
the effect of $u$ and $d$ sea quarks~\cite{qcdsf, etmpiff, Brandt:2013dua} 
to those with only a single 
lattice spacing including the effect of $u$, $d$ sea 
quarks~\cite{Aoki:2009qn} or, more realistically, 
$u$, $d$ and $s$ sea 
quarks~\cite{Aoki:2015pba, Nguyen:2011ek, rbcukqcd}. 
See also~\cite{Fukaya:2014jka} for a calculation 
in the $\varepsilon$ regime. A mean square radius can 
similarly be defined for the scalar form factor. 
Earlier results, again for 
relatively heavy values of the $\pi$ meson mass, have 
been obtained with $u$ and $d$ sea quarks (only) 
in~\cite{Aoki:2009qn, Gulpers, Gulpers:2015bba}.   

Here we give results for both vector and scalar form factors for 
$\pi$ mesons made of physical $u/d$ quarks and including 
a fully realistic quark content in the sea, with
physical $u$, $d$, $s$ and $c$ quarks. We also work with 
three different values of the lattice spacing. This enables 
good control of systematic errors both from $m_{\pi}$ and 
from discretisation. Our lattices have large volumes 
with a minimum spatial size of 4.8 fm. 

The vector form factor is accessible in experiment 
and, as we shall see, our results can be directly compared 
to the experimental data with no extrapolation. 

The scalar form factor cannot be obtained directly 
by experiment
but information on it can be extracted by 
applying chiral perturbation 
theory to the $\pi$ decay constant and 
to $\pi-\pi$ scattering~\cite{gasserl, Gasser:1983kx, Colangelo:2001df}.
An additional
ingredient in the lattice QCD calculation in 
this case is the need to include 
quark-line disconnected contributions. 
The expectation from chiral perturbation 
theory~\cite{Juttner:2011ur} is for
the disconnected contribution to the form factor 
at $q^2=0$ to be small but for the impact on 
the radius as defined in eq.~(\ref{eq:rdef}) to
be substantial. 
Our results are very much in line with 
expectations from chiral perturbation theory 
and we are able to distinguish disconnected 
contributions coming from $u/d$ and $s$ quark 
loops. 

Section~\ref{sec:latt} describes how the lattice calculation 
is done and gives details of the results. 
Our results are compared to experiment, to chiral perturbation 
theory expectations, and to other lattice 
calculations in Section~\ref{sec:discussion} and 
Section~\ref{sec:conclusions} gives our conclusions, 
looking forward to improved calculations in future. 

\section{Lattice Calculation} 
\label{sec:latt}

For the lattice QCD calculation we use the Highly Improved Staggered 
Quark (HISQ) action~\cite{HISQ_PRD}, which has been demonstrated 
to have very small discretisation errors~\cite{jpsi, fkpi}. 
We use gluon field configurations generated by the 
MILC collaboration~\cite{Bazavov:2010ru, Bazavov:2012xda} 
that include $u$, $d$, $s$ and $c$ sea 
quarks using the HISQ action along with a fully 
$\mathcal{O}(\alpha_sa^2)$ improved gluon action~\cite{Hart:2008sq}. 
The ensembles that we use here have light quark masses 
$m_u=m_d=m_{l}$ with $m_l$ and hence $m_{\pi}$ close to its physical value. 
The parameters of the ensembles are given in Table~\ref{tab:params}.  

\begin{table}
\caption{
The MILC gluon field ensembles (sets) used 
here~\cite{Bazavov:2010ru, Bazavov:2012xda}. 
The lattice spacing, $a$, is determined using the 
$w_0$ parameter~\cite{Borsanyi:2012zs}, and has a correlated 
0.5\% uncertainty from the physical value of 
$w_0$, fixed using $f_{\pi}$~\cite{fkpi}. 
Set 1 will be referred to ``very coarse'', 
2 as ``coarse'' and 3 as ``fine''. 
Columns 3, 4 and 5 give the sea quark masses 
in lattice units 
($m_u=m_d=m_{l}$).
$L_s$ and $L_t$ 
are the lengths in lattice units in space and 
time directions for each lattice. 
The number of 
configurations that we have used in each set is given 
in the seventh column. 
The final column gives the values of the end-point 
of the 3-point function, $T$, in lattice units.  
}
\label{tab:params}
\begin{ruledtabular}
\begin{tabular}{lllllllll}
Set & $a$/fm &  $am_{l,sea}$ & $am_{s,sea}$ & $am_{c,sea}$ & $L_s\times L_t$ & $n_{\mathrm{cfg}}$ & $T$ \\
\hline
1 & 0.1509 &  0.00235  & 0.0647  & 0.831 & 32$\times$48 & 1000 & 9,12,15 \\
\hline
2 & 0.1212 &  0.00184 & 0.0507 & 0.628 & 48$\times$64 & 1000 & 12,15,18 \\
\hline
3 & 0.0879 &  0.0012 & 0.0363 & 0.432 & 64$\times$96 & 223 & 16,21,26 \\
\end{tabular}
\end{ruledtabular}
\end{table}

\begin{figure}
\centering
\includegraphics[width=0.3\textwidth]{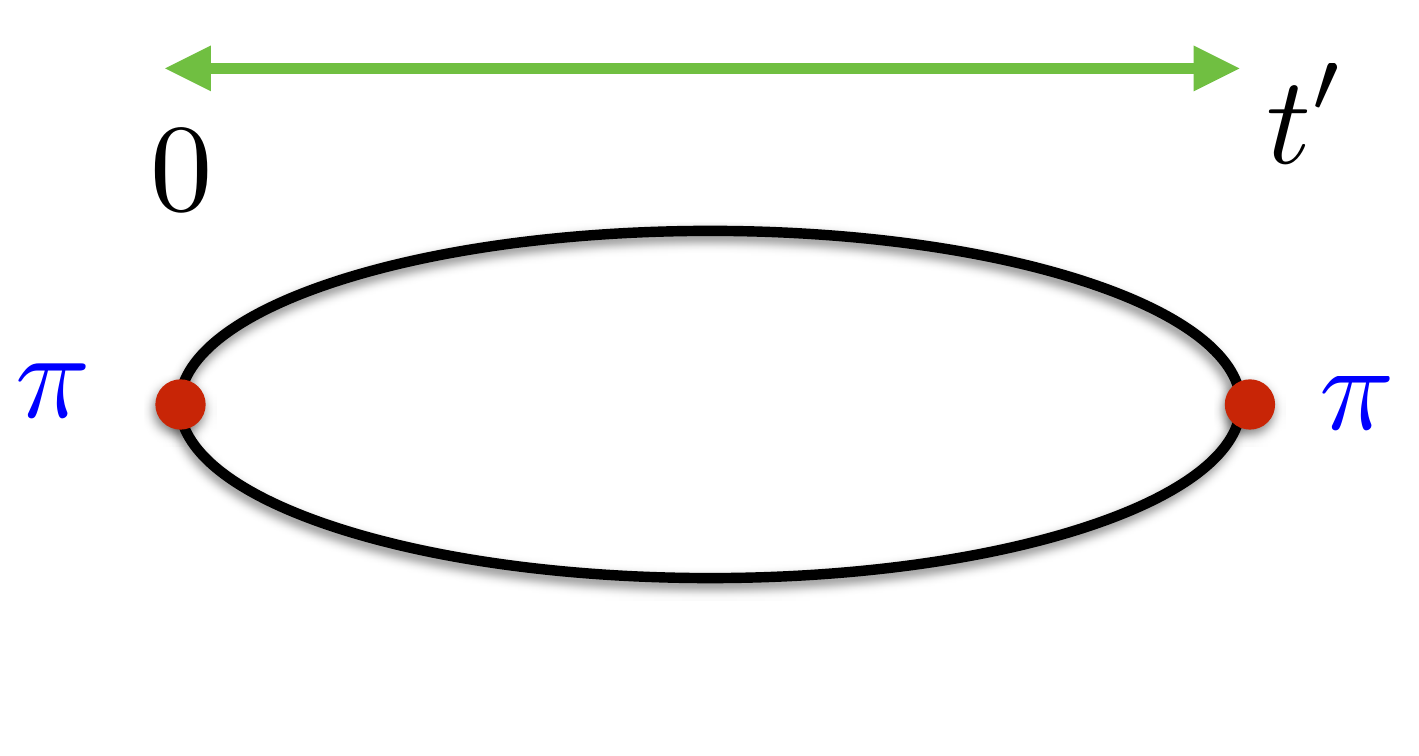}
\vspace{3mm}
\includegraphics[width=0.3\textwidth]{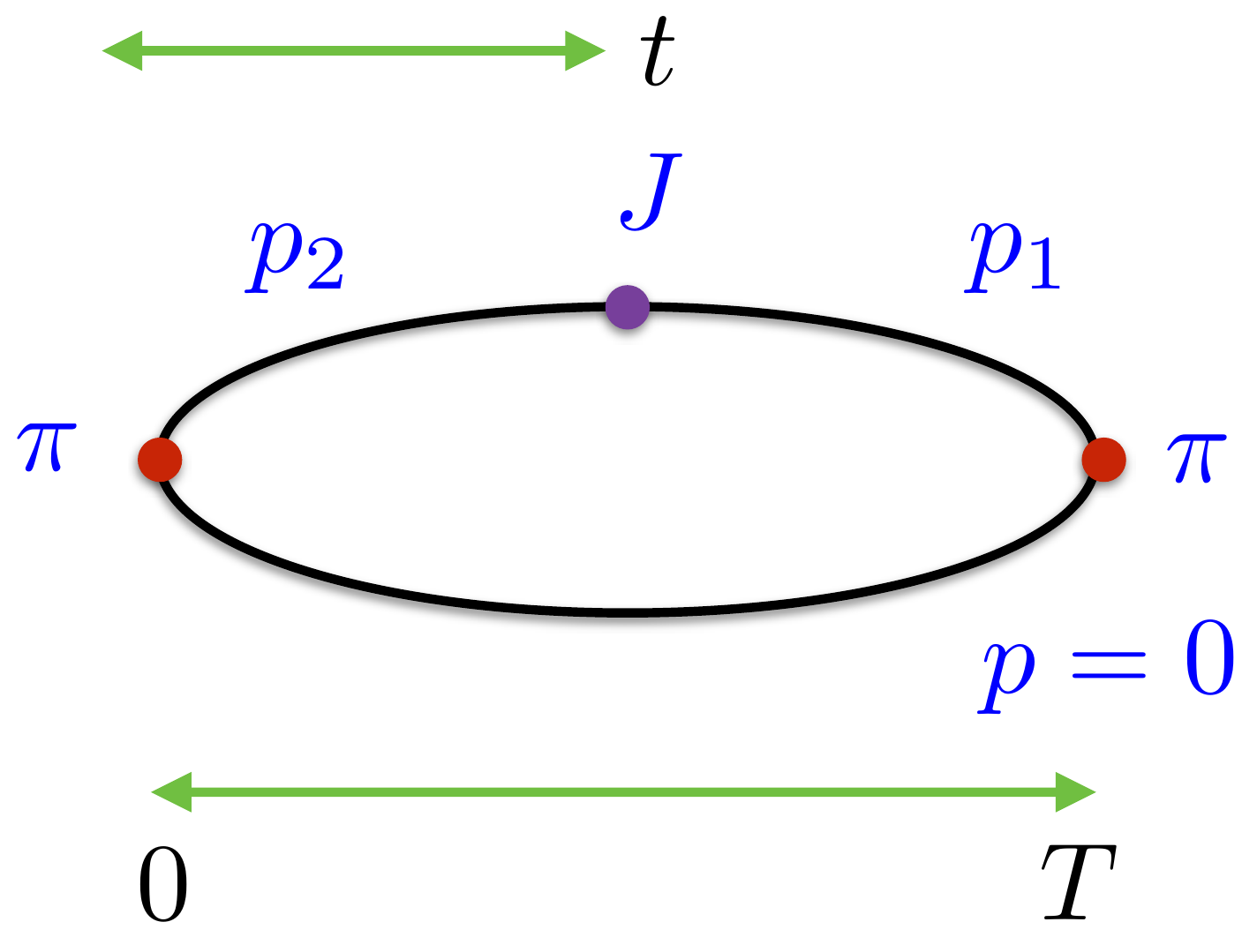}
\includegraphics[width=0.3\textwidth]{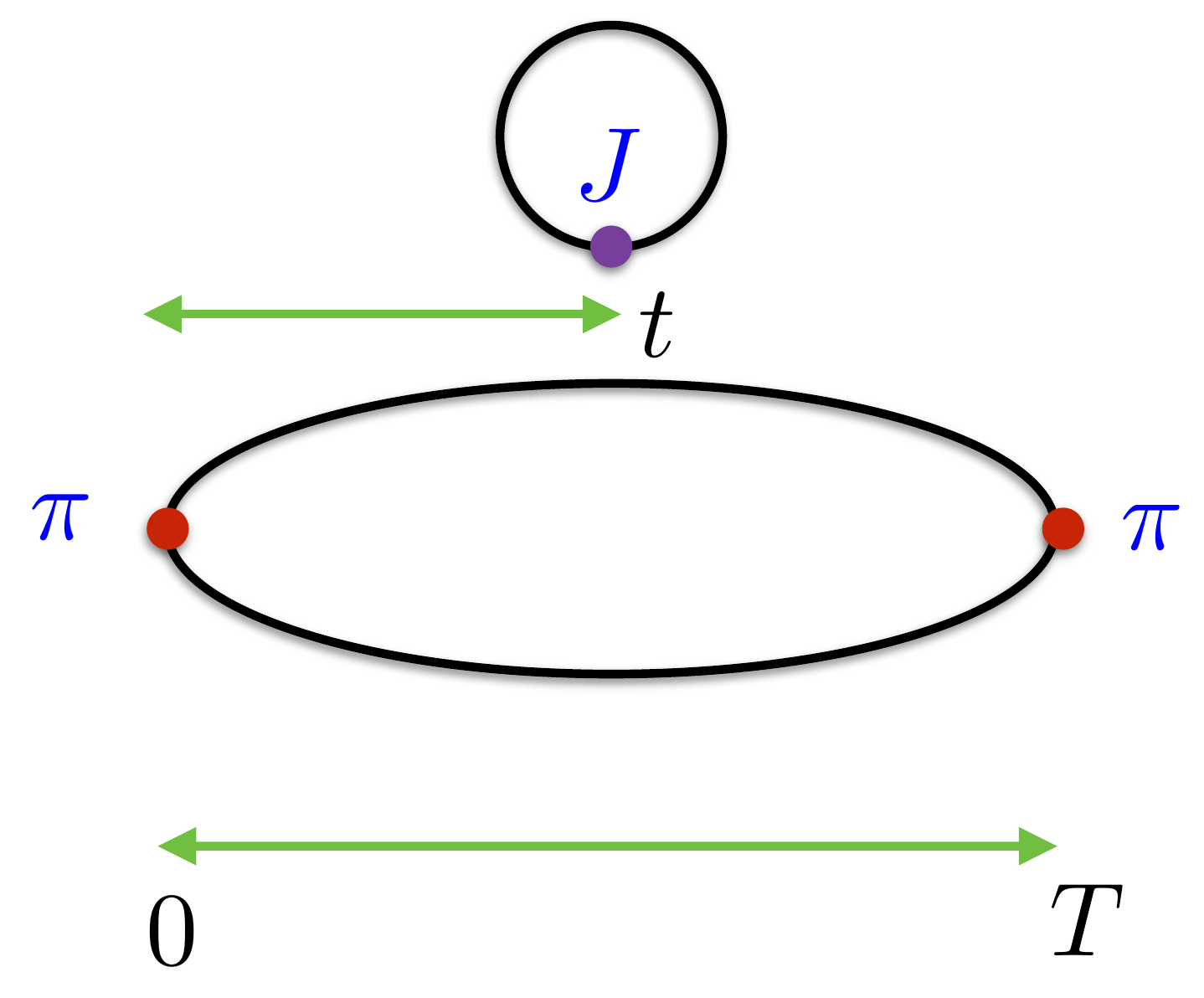}
\caption{2-point (top) and 3-point quark-line-connected (middle) 
and quark-line disconnected (bottom) correlators.}
\label{fig:2pt3ptdiagram}
\end{figure}

On these configurations we generate HISQ light quark propagators 
with the same mass as that of the  
sea light quarks. We use a local random wall source~\cite{fkpi} and 
4 time sources per configuration for high statistics. 
The propagators are combined into $\pi$ meson correlation functions 
(2-point correlators) that create a $\pi$ meson at time 0 and 
destroy it at time $t^{\prime}$ and correlation functions that allow 
for interaction with a current $J$ at an intermediate time, $t$,  
between a $\pi$ meson source at 0 and sink at $T$ (3-point correlators). 
These are 
illustrated in Fig.~\ref{fig:2pt3ptdiagram}. 
Results at all $t^{\prime}(t)$ values are obtained for 2(3)-point functions 
and we also use three 
values for $T$ in the 3-point functions, 
so that our fits can map out fully the $t$ and $T$ 
dependence for improved accuracy. 
When $J$ is a vector current we need to consider only 
one 3-point diagram for the flavour non-singlet $\pi$.
This is shown as the central diagram of Fig.~\ref{fig:2pt3ptdiagram} 
in which the current $J$ is inserted into one of the legs
of the 2-point function. We simply multiply by 2 to allow 
for its insertion into the other leg. 
The `disconnected diagram' which is the product of a 
$\pi$ 2-point function and a closed quark loop coupled 
to $J$ is shown as the lower diagram of Fig.~\ref{fig:2pt3ptdiagram}. 
This vanishes for vector $J$ in the ensemble average because it is odd under 
charge-conjugation~\cite{Draper:1988bp}.
For scalar $J$ this diagram needs to be included and 
different combinations of flavours of quarks in the closed 
quark loop give rise to different form factors. 

The $\pi$ mesons in our correlators are the Goldstone mesons 
whose mass vanishes with $m_{l}$. We ensure this by 
using the local $\gamma_5$ operator at source and sink. In 
staggered quark parlance this is the $\gamma_5 \otimes \gamma_5$ operator. 
For $J$ we use a symmetric 1-link point-split spatial vector 
current, $V_i$, or a local scalar current, $S$. A gluon field is included 
in the vector current to make it gauge-covariant. Both of these are `tasteless' 
staggered quark operators ($\gamma_i \otimes 1$ and $1 \otimes 1$) 
and so can be used in a 3-point function with the Goldstone meson 
at source and sink. 

We work with several $\pi$ meson spatial momenta 
by generating 
light quark propagators with a phase included 
on the spatial gluon links. This is equivalent to 
introducing a phase into the boundary condition on the 
field~\cite{twist}, which gives a momentum 
to the quark. This is referred to as using `twisted 
boundary conditions'. As illustrated in the central 
diagram of Fig.~\ref{fig:2pt3ptdiagram} 
we choose the spectator quark in the 3-point function to have zero momentum 
and give momenta $p_1$ and $p_2$ to the quarks that 
interact with the current. 
Both momenta are chosen to be in the $(1,1,1)$ 
direction. By using various values of $p_1$ and $p_2$ we 
can obtain 3-point functions at several different 
small values of squared 4-momentum transfer, $q^2$. 

\subsection{Vector Form Factor}
\label{sec:vec}

\begin{figure}
\centering
\includegraphics{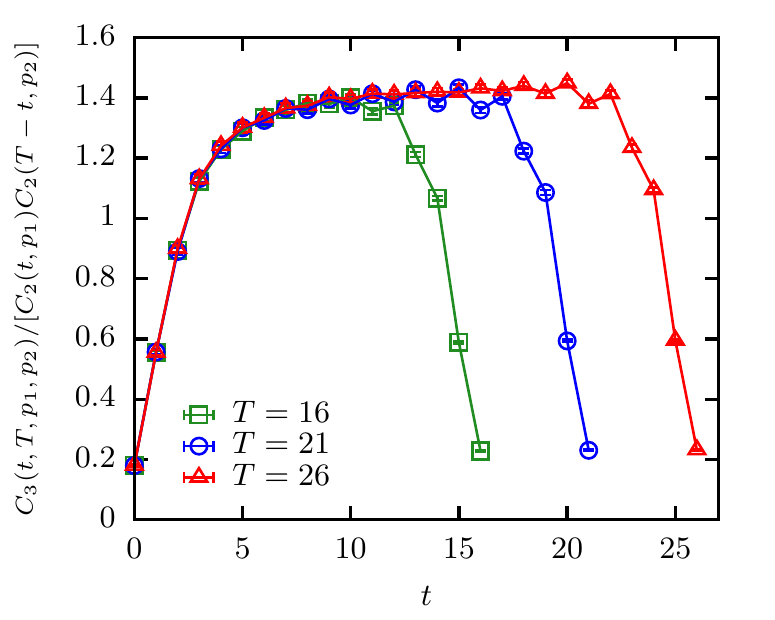}
\caption{Results for the ratio of the 3-point correlator containing 
a vector current to the 
product of appropriate 2-point correlators for the pion on 
fine ensemble set 3. 
The ratio for the 3 different values of $T$ 
are plotted as a function of $t$ with a 
$\pi$ of momentum zero on the left, and momentum $ap = 0.0363$ 
on the right. Note that this figure is to illustrate the 
quality of our results; we do {\it not} use this ratio 
to extract ground-state parameters. Instead we perform a simultaneous 
fit to multiple exponentials for both the 2-point and 3-point 
correlators as described in the text.  
}
\label{fig:vec2pt3pt}
\end{figure}

For the vector current case we have a set of 2-point and 3-point quark-line 
connected correlators at various values of $p_1$ and $p_2$ on each 
ensemble.  
The quality of our results is illustrated for one 
ensemble and set of momenta in Figure~\ref{fig:vec2pt3pt}. 
The 2-point and 3-point correlators are all fit 
simultaneously using Bayesian methods~\cite{gplbayes} 
that allow us to include the effect of excited states, both 
`radial' excitations and, because we are using staggered quarks, 
opposite parity mesons that give oscillating terms. Since the oscillating 
terms are absent for zero-momentum $\pi$ mesons they are small 
here, but we nevertheless include them in our fits. Having 
3-point correlators from multiple $T$ values is also important 
in taking account of excited states. Fitting multiple 
momenta simultaneously allows us to take account of correlations 
between the correlators. 

The fit form for the 
2-point function with source at 0 and sink at $t^{\prime}$ and spatial 
momentum, $p$, is: 
\begin{eqnarray}
C_{2pt}(p,0,t^{\prime}) &=& \sum_{i} b_{i}^2(p) \mathrm{fn}(E_{i}(p),t^{\prime}) + \mathrm{o.p.t.} \nonumber \\
\mathrm{fn}(E,t) &=& e^{-Et} + e^{-E(L_t-t)}. 
\label{eq:2ptfit}
\end{eqnarray}
Opposite parity terms (o.p.t.) are similar to the terms given explicitly above 
but with factors of $(-1)^{t^{\prime}/a}$. 
For the 3-point function~\cite{Koponen:2013tua, Donald:2013pea}:
\begin{multline}
C_{3pt}(p_1,p_2,0,t,T) = \sum_{i,j} \big [b_{i}(p_1) \mathrm{fn}(E_{i}(p_1),t) \times \\
J_{i,j}(p_1,p_2) b_{j}(p_2) \mathrm{fn}(E_{j}(p_2),T-t)\big ] + \mathrm{o.p.t.}  
\label{eq:3ptfit}
\end{multline}
Prior values and widths are taken as: ground-state energy, 10\% width; splitting 
between ground-state and excited energies, 650 MeV with 50\% width; splitting 
between ground-state and lowest oscillating state, 500 MeV with 50\% width;
amplitudes, 0.01(1.0) for normal states 
and 0.01(0.5) for oscillating states; matrix elements, 0.01(1.0) for vector
currents. 
 We take the result from a 6 exponential fit (with 6 oscillating exponentials)
to obtain the vector form factor.  

\begin{table*}
\caption{
Upper table: Results for $\pi$ energies in lattice units
at the different spatial momenta used on each set, as well as the corresponding 
amplitudes from the 2-point functions. The values given here come from 
the simultaneous fit of 2-point correlators with 3-point correlators 
containing a vector current. Results for 2-point 
parameters from the fit of 2-point correlators 
with 3-point correlators including a scalar current are in agreement.  
$\pi$ results at zero momentum agree with those in~\cite{fkpi,cond}, 
but are not the same because the fits used here also 
include 3-point functions. \\
Lower table: Results for unrenormalised form factors at $q^2$ values corresponding 
to different combinations of $\pi$ momenta (from the upper table) at 
source and sink. The results 
at $q^2=0$ come from using the lowest non-zero spatial momentum for both
$p_1$ and $p_2$. The scalar form factor results given are for the 
connected 3-point function only.  
}
\label{tab:mep}
\begin{ruledtabular}
\begin{tabular}{llllllllll}
Set & $pa$ &  $aE_{\pi}$ & $a^{3/2}b$ & $pa$ & $aE_{\pi}$ & $a^{3/2}b$ & $pa$ & $aE_{\pi}$ & $a^{3/2}b$\\
\hline
1 & 0.0 &  0.10167(5)  & 0.4845(3) & 0.0623  & 0.11921(6) & 0.4465(2) & 0.2490 & 0.2669(9) & 0.2936(14)\\
\hline
2 & 0.0 &  0.08159(3) & 0.35773(15) & 0.05 & 0.09569(4) & 0.32981(14) & 0.16482 & 0.1840(2) & 0.2375(3)\\
  & 0.2 & 0.2161(4) & 0.2193(5)  & & & & & & \\
\hline
3 & 0.0 &  0.05720(3) & 0.23272(15) & 0.0363 & 0.06767(4) & 0.21397(13) & 0.1451 & 0.1546(5) & 0.1400(5)\\
\hline
\hline
Set & $q^2a^2$ & $f_+(q^2)/Z_V$ & $f_0^{\mathrm{conn}}(q^2)/Z_S$ & $q^2a^2$ & $f_+(q^2)/Z_V$ & $f_0^{\mathrm{conn}}(q^2)/Z_S$ & $q^2a^2$ & $f_+(q^2)/Z_V$ & $f_0^{\mathrm{conn}}(q^2)/Z_S$ \\  
\hline
1 & 0.0 & 0.837(3) & 2.163(6) & -0.0036 & 0.832(4) & 2.143(4) & -0.0346 & 0.761(8) & 1.98(2) \\
 & -0.0751 & 0.678(10) & 1.82(2) &  &  & &  &  &  \\
\hline
2 & 0.0 & 0.852(2) & 1.769(3) & -0.0023 & 0.847(3) & 1.753(2) & -0.0054 & 0.838(4) & 1.719(6) \\
 & -0.0167 & 0.797(3) & 1.656(5) & -0.0220 & 0.782(4) & 1.623(7) & -0.0384 & 0.731(5) & 1.542(7) \\
 & -0.0480 & 0.702(8) & 1.500(8) &  &  & &  &  &  \\
\hline
3 & 0.0 & 0.841(2) & 1.330(4) & -0.0012 & 0.842(4) & 1.321(3) & -0.0116 & 0.775(7) & 1.210(10) \\
 & -0.0254 & 0.692(8) & 1.125(10) &  &  & &  &  & \\
\end{tabular}
\end{ruledtabular}
\end{table*}

The ground-state parameters are given by $i=j=0$ in eqs.~(\ref{eq:2ptfit}) 
and~(\ref{eq:3ptfit}) and are our key results. 
By matching to a continuum 
correlator with a relativistic normalisation of states and allowing 
for a renormalisation of the lattice current, we see that the 
matrix elements between the ground state mesons that we want to determine
are given by: 
\begin{equation}
\langle \pi(p_1) |J| \pi(p_2) \rangle = Z\sqrt{4E_{0}(p_1)E_{0}(p_2)}J_{0,0}(p_1,p_2).
\label{eq:me}
\end{equation}
The matrix element is related to the form factor for the vector current via: 
\begin{equation}
\langle \pi(p_1) |V_i| \pi(p_2) \rangle = f_+(q^2)(p_1+p_2)_i 
\end{equation}
The vector matrix element can be normalised using the fact that 
$f_+(0)=1$ for a conserved current (inserted in either the quark 
or the antiquark legs in Fig.~\ref{fig:2pt3ptdiagram}), and we can therefore determine 
$Z$ by demanding that condition for our current. 
$f_+(0)$ is determined at $q^2=0$ for spatial $V_i$ by setting $p_1=p_2 \neq 0$. 

\subsubsection{Results} 

Table~\ref{tab:mep} gives results for the $\pi$ energies and 
2-point amplitudes as a function of momentum. A good test of 
discretisation errors is to determine the speed of light, $c^2$ 
from $(E^2-m^2)/p^2$. From Table~\ref{tab:mep} we see that 
$c^2$ deviates from 1 at most by 2(1)\% at the largest momenta.  
Another test is to compare the scaling of amplitudes to the 
expected $1/\sqrt{E}$ behaviour for a pseudoscalar. 
Again we see good agreement, with deviations at most 3(1.5)\%. 

\begin{figure}
\centering
\includegraphics{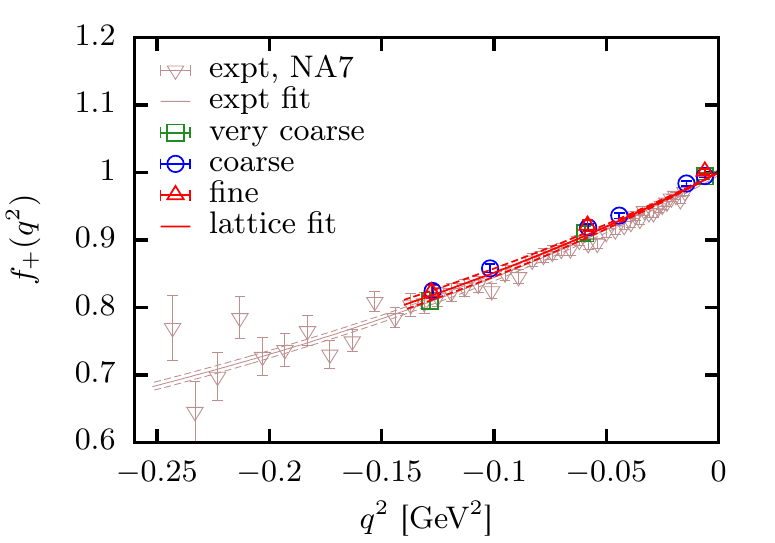}
\caption{Lattice QCD results for the vector form factor 
on each ensemble compared directly to the  
experimental results from~\cite{amendolia}. 
Fit curves for both experiment and lattice QCD results 
are given to a `monopole' form. 
}
\label{fig:fvsexp}
\end{figure}

Table~\ref{tab:mep} also gives the raw (unrenormalised) form 
factors for various $q^2$ values obtained from 
different combinations of momenta (in positive and 
negative) directions at source and sink. The statistical 
errors on the form factors are 0.5-3\%. By dividing the 
values at non-zero $q^2$ by the value at $q^2=0$ we obtain 
normalised values for $f_+$. $f_+$ is plotted against
$q^2$ in Fig.~\ref{fig:fvsexp} for all three sets along 
with the results from experiment~\cite{amendolia}. 
The agreement with experiment is good, 
reflecting the fact that our results correspond to physical 
$\pi$ masses.  

In fitting a functional form in $q^2$ to our 
results to extract a mean squared radius, 
we use the same form as that used for the 
experimental results~\cite{amendolia}, 
but including allowance for finite lattice spacing effects. 
We also fit over a similar range of $q^2$ values. 
We use:
\begin{equation}
\label{eq:rfit}
f(q^2)=\frac{1}{1+c_{a^2}(\Lambda a)^2+c_{a^4}(\Lambda a)^4-\langle r^2\rangle q^2/6}
\end{equation}
(note that $q^2$ is negative), where
\begin{eqnarray}
\label{eq:rform}
&&\langle r^2\rangle_V(a,\delta m_{\mathrm{sea}}, m_{\pi}) = \\
&&\langle r^2\rangle_{V}\bigg[1 + b_{a^2}(\Lambda a)^2 + b_{a^4}(\Lambda a)^4+\frac{b_{\mathrm{sea}}\delta m_{\mathrm{sea}}}{10m_{s,\mathrm{phys}}}\bigg] \nonumber \\
&-&\frac{1}{\Lambda_{\chi}^2}\ln\bigg(\frac{m^2_{\pi}}{m^2_{\pi,\mathrm{phys}}}\bigg). \nonumber
\end{eqnarray}
Here $c_{a^n}$ and $b_{a^n}$ allow for discretisation effects in 
the normalisation of $f_+$ and in $\langle r^2 \rangle$ respectively. 
We take $\Lambda$ = 500 MeV and allow priors on the $b$ and $c$ fit 
parameters of 0.0(1.0) for $b_{a^4}$ and $c_{a^4}$ and 
0.0(0.3) on $b_{a^2}$ and $c_{a^2}$ (since tree-level $a^2$ 
errors are absent in this calculation). We allow a prior width 
on the physical result for the mean squared radius, $\langle r^2 \rangle_{V}$ 
of 25\%.  
The term with coefficient $b_{\mathrm{sea}}$ allows for mistuning
of sea quark masses. From chiral perturbation theory a term linear in the 
quark masses is expected, and it is convenient to take this term 
as a ratio to another quark mass so that factors of the quark 
mass renormalisation cancel. The factor of 10 multiplying $m_{s,\mathrm{phys}}$ 
gives a value close to the chiral scale, $\Lambda_{\chi}$. 
The mistuning of the 
sea masses is defined as $\sum_{u,d,s}(m_q-m_q^{\mathrm{tuned}})$ 
and values of $\delta m_{\mathrm{sea}}/m_{s,\mathrm{phys}}$ 
values for these ensembles are tabulated in~\cite{Chakraborty:2014aca}. 
The values are all less than 0.05, but not zero because of 
mistuning of the sea $s$ quark mass. 
 
The final logarithmic term in eq.~(\ref{eq:rform}) 
comes from chiral perturbation theory~\cite{gasserl} and 
is the source of the divergence in the radius as $m_{\pi} \rightarrow 0$. 
We use it, rescaling the argument of the 
logarithm so that it vanishes at the physical pion 
mass, to make small adjustments for the fact that our 
$u/d$ quark masses are not exactly at their physical values 
(in fact they are slightly too low). $\Lambda_{\chi}$ = 1.16 GeV. 
Because we are very close to 
the physical light quark mass point we do not need to include 
further terms in a chiral perturbation theory expansion since 
they will be negligible. 

We apply the functional form of eq.~(\ref{eq:rfit}) and~(\ref{eq:rform}) 
to our result taking account of the correlations between results at 
different values of $q^2$ obtained on a given ensemble.
The fit has $\chi^2/\mathrm{dof} = 0.9$ and gives the physical 
result for the electric charge radius of the 
$\pi$ of $\langle r^2 \rangle_{V} =$ 10.35(46) ${\mathrm{GeV}}^{-2}$, 
or 0.403(18) $\mathrm{fm}^2$. 

We can also use the final logarithmic term in eq.~(\ref{eq:rform}) to 
estimate the impact of isospin and electromagnetic 
effects by varying the value of $m_{\pi,\mathrm{phys}}$ used 
there. The physical value of $m_{\pi}$ corresponding to our  
lattice world in which $u$ and $d$ quark masses are equal and there 
is no electromagnetism is $m_{\pi^0}$ = 0.135 GeV~\cite{Aubin:2004fs}, 
and we use this 
for our central value above. The experimental results correspond 
to $m_{\pi^+}$ = 0.139 GeV and we substitute that for the 
physical value in the logarithm to assess the uncertainty 
from the fact that the real world has different $u/d$ quark 
masses and the quarks have electric charge. 
This gives an estimate for the systematic uncertainty 
from isospin/electromagnetism 
of 0.5\%. 

We must also include a systematic uncertainty from 
working on lattices with finite spatial volume, albeit large. 
Finite volume effects are small on these lattices for the 
$\pi$ mass and decay constant~\cite{fkpi} and effects 
of similar size are expected in the form factor at 
fixed $q^2$.  
Because the mean squared radius is defined from the small 
difference in values for the form factor as 
$q^2$ moves away from zero (where the form factor is defined to be 1), 
a small effect on the form factor at non-zero $q^2$ can 
become a significant effect on the radius. 
These effects can be estimated from chiral 
perturbation theory. Continuum chiral perturbation theory 
is a good guide here and we do not need staggered chiral perturbation 
theory because, as shown in~\cite{Colquhoun:2015mfa}, 
staggered quark taste-effects which might be expected to affect 
$\pi$ masses appearing in chiral loops in fact tend to cancel 
against associated hairpin diagrams. It turns out that this 
cancellation happens for a wide range of quantities (including 
decay constants and form factors) for a specific value of the 
hairpin coefficients that seems to be close to the value obtained 
in practice. We therefore use continuum analyses and specifically results from 
analyses that are relevant to our use of twisted boundary
conditions~\cite{Jiang:2006gna, Bijnens:2014yya} because this 
modifies the expected finite-volume dependence.
From~\cite{Jiang:2006gna} the relative finite volume effect in 
the vector squared-radius varies in the range 1--1.5\% 
for lattice sizes that we use in the range 4.8 fm to 5.8 fm
for physical $\pi$ masses. 
Note that the direction of the finite-volume 
effect is such that the radius would be larger in the infinite volume 
limit.  
We do not make a correction for this but include an uncertainty of 1.5\% 
for finite volume effects. 

\begin{table}
\caption{
Error budget for the mean square radii of the $\pi$,  
as a percentage of the final answer. See the discussion 
in the text for a description of each component.  
}
\label{tab:error}
\begin{ruledtabular}
\begin{tabular}{llll}
 & $\langle r^2 \rangle_V$ &  $\langle r^2 \rangle_S^{\mathrm{conn}}$ & $\langle r^2 \rangle_S^{\mathrm{singlet/octet}}$ \\
\hline
statistics/fitting & 4.5 & 5   &  7.5/8.5 \\
isospin/electromagnetism  & 0.5 & 3  & 3 \\
finite volume & 1.5 & 10 & 10 \\
total & 4.8 & 12 & 13/14
\end{tabular}
\end{ruledtabular}
\end{table}

Our error budget for $\langle r^2 \rangle_V$ is given 
in Table~\ref{tab:error}. Adding the systematic uncertainties in 
quadrature as the second uncertainty gives our result: 
\begin{equation}
\label{eq:rvres}
\langle r^2 \rangle_V^{(\pi)} = 0.403(18)(6) \,{\mathrm{fm}}^2  
\end{equation}
to be compared to 0.431(10) $\mathrm{fm}^2$ from the 
experimental results of~\cite{amendolia} using the same 
fit form. The Particle Data Group~\cite{PDG2014} give a mean square 
radius from averaging over several experimental results  
of 0.452(11) $\mathrm{fm}^2$. 

\subsection{Scalar Form Factor} 
\label{sec:scalar}

\subsubsection{Results for the connected contribution}
\label{sec:connscalar}

We begin by discussing our results for the connected 
contribution to the scalar form factor of the $\pi$. 
This is the result calculated from 3-point functions of 
the form sketched in the central diagram of 
Fig.~\ref{fig:2pt3ptdiagram} in which the scalar 
current is composed of the light quarks which are the 
valence quarks of the $\pi$. 
Although this form factor does not correspond to a physically 
realisable process (even if we had a particle with 
which to produce a scalar current) 
it is nevertheless possible and useful to compare 
different lattice QCD calculations for it. 
Different formalisms within lattice QCD should give 
the same results in the continuum and chiral limits 
for the mean square radius from the connected scalar 
form factor. A key issue, to be discussed further below, 
is then how big the additional contribution is from 
quark-line disconnected diagrams. 

\begin{figure}
\centering
\includegraphics{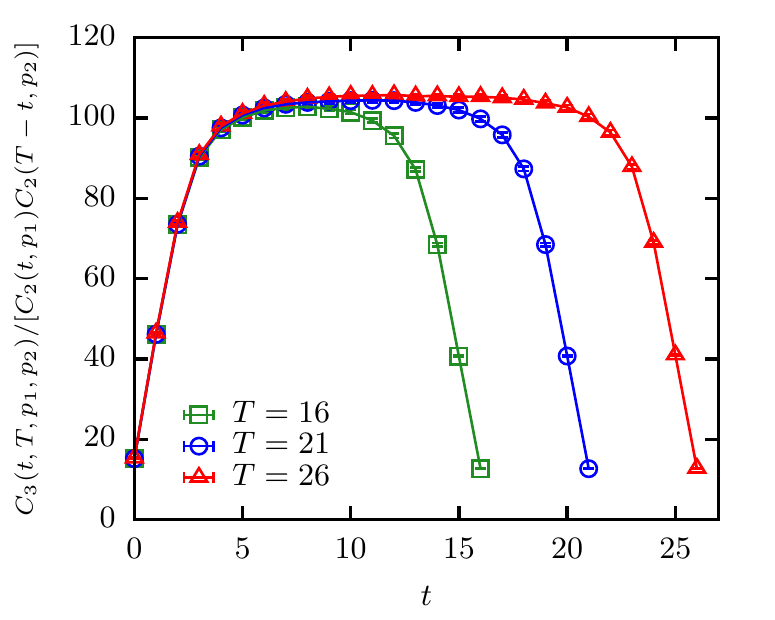}
\caption{Results for the ratio of the 3-point connected correlator containing 
a scalar current to the 
product of appropriate 2-point correlators for the pion on 
fine ensemble set 3. 
The ratio for the 3 different values of $T$ 
are plotted as a function of $t$ with a 
$\pi$ of momentum zero on the left, and momentum $ap = 0.0363$ 
on the right. Note that this figure is to illustrate the 
quality of our results; we do {\it not} use this ratio 
to extract ground-state parameters. Instead we perform a simultaneous 
fit to multiple exponentials for both the 2-point and 3-point 
correlators as described in the text.  
}
\label{fig:s2pt3pt}
\end{figure}

\begin{figure}
\centering
\includegraphics{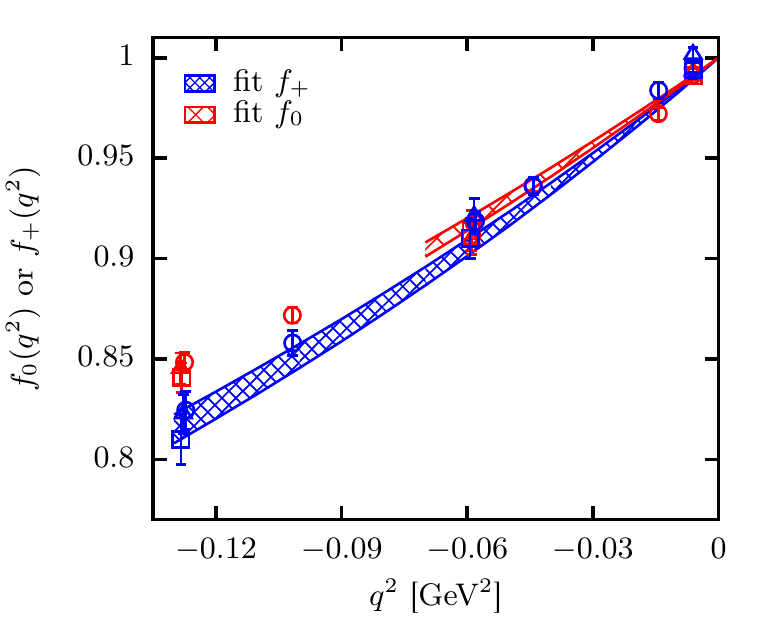}
\caption{Comparison of our lattice QCD results for the pion 
vector form factor (blue) with the connected part of the pion scalar
form factor (red). Results from set 1 are shown as open
squares, set 2 as circles and set 3 as triangles. The hashed curves give 
the fit to the form factors described in the text (using a smaller 
$q^2$ range in the scalar case than the vector). 
}
\label{fig:fpf0}
\end{figure}

The calculation for the connected scalar form factor proceeds 
in an identical way to that of the vector form factor 
discussed in Section~\ref{sec:vec}. 
We calculate the 3-point function given as the central 
figure of Fig.~\ref{fig:2pt3ptdiagram} with a scalar 
current made from light quarks inserted as $J$. We use the same light quark 
propagators and 2-point functions as for the vector case. 
The quality of our results is illustrated for one 
ensemble and set of momenta in Fig.~\ref{fig:s2pt3pt} 
(we use the same ensemble and set of momenta as in Fig.~\ref{fig:vec2pt3pt}). 

We fit the 2-point and 3-point correlators simultaneously 
(but in a separate fit from the vector case)
as a function of $t$, $t^{\prime}$ and $T$ as given 
in eqs.~(\ref{eq:2ptfit}) and~(\ref{eq:3ptfit}). 
The priors are taken to be the same as in the vector 
case except that the prior width on the scalar matrix 
element is taken to be much larger, reflecting expectations 
on its value given below. We take the prior width 
on the scalar matrix element to be 20.0 on the very coarse 
ensemble and 25.0 on the coarse and fine ensembles. 

The ground-state matrix element for the scalar current is 
related to the parameter $J_{0,0}$ extracted from 
our fits as in eq.~(\ref{eq:me}).  In turn the 
matrix element is related to the form factor that 
we wish to extract by  
\begin{equation}
\label{eq:sff}
\langle \pi(p_1) |S| \pi(p_2) \rangle^{\mathrm{conn}} = Af_0^{\mathrm{conn}}(q^2)
\end{equation} 
where $A$ is a normalisation factor. 
Our scalar current made from HISQ quarks 
is absolutely normalised~\cite{Na:2010uf}. If we had included 
disconnected diagrams associated with the scalar current we would be 
able to write, from the Feynman-Hellmann theorem, 
\begin{equation}
\label{eq:sff-fh}
\langle \pi(p_1) |S| \pi(p_2) \rangle = f_0(q^2)\frac{\partial m_{\pi}^2}{2\partial m_{l}} , 
\end{equation} 
with $f_0(0)=1$, $A=(\partial m_{\pi}^2/\partial m_l)/2$ and 
we take the same $m_l$ value for valence and sea $l$ quarks. 
The factor of 2 on the right-hand side comes from the fact that we 
are inserting a scalar current in only one propagator to make the 
3-point correlator and the $\pi$ has two valence light quarks. 
For the connected correlator we expect instead that factor $A$ in 
eq.~(\ref{eq:sff}) should be 
equal to half the derivative of the squared $\pi$ mass with respect to 
its valence quark mass. This can be tested approximately using
$\pi$ and $K$ masses on these ensembles in~\cite{fkpi}. 
Comparing $a(m_K^2-m_{\pi}^2)/(m_s-m_l)$ (i.e. an approx 
derivative for a pseudoscalar meson mass) to the result for 
the unrenormalised 
$f_0^{\mathrm{conn}}(0)/Z_S$ (=$A$) from Table~\ref{tab:mep} shows 
agreement within 10\%, confirming that $A$ does have the 
expected value.  
Since here we are chiefly concerned with the shape of the 
form factor, we simply treat the scalar current as requiring a 
$Z$ factor, $Z_S$, 
and determine this from 
the requirement that also $f_0(0)=1$.  

To determine the mean squared radius associated with the 
connected scalar form factor we take the same fit as for 
the vector case, eqs.~(\ref{eq:rfit}) and~(\ref{eq:rform}), 
except that the coefficient of the chiral logarithm is 
now 6. This coefficient applies to the complete scalar 
form factor but we use it here to estimate conservatively the 
impact of changing the $\pi$ mass close to the physical point 
and therefore the uncertainty. 
We use the same priors on the coefficients as in 
the vector fit, except that we increase the prior on 
the physical result for the mean squared radius, since its 
value is less well-known.  

Our fit has a $\chi^2/{\mathrm{dof}}$ of 1.1 and gives a 
physical result, $\langle r^2 \rangle_{S}^{\mathrm{conn}}$ of 
8.97(45) $\mathrm{GeV}^{-2}$ or 0.349(18) $\mathrm{fm}^2$. 
The systematic errors are somewhat larger in the scalar 
case because of the larger coefficient of the chiral 
logarithm. Using this we estimate the systematic uncertainty 
from isospin/electromagnetism at 3\%.
The larger coefficient for the chiral logarithm also carries 
with it the implication of larger finite volume effects, 
potentially by a factor of 6, giving a systematic uncertainty 
of 10\% on our ensembles from this source, allowing for the 
fact that the mean square radius is slightly smaller than 
for the vector case. 

Adding systematic errors in quadrature our final 
result for the mean squared radius from the 
connected scalar form factor is 
\begin{equation}
\label{eq:rsres}
\langle r^2 \rangle_{S,\mathrm{conn}}^{(\pi)} = 0.349(18)(36) {\mathrm{fm}}^2  
\end{equation}
Our error budget is given 
in Table~\ref{tab:error}. 
This radius has a central value that is only slightly smaller 
than the vector form factor radius (eq.~(\ref{eq:rvres})). 
This is illustrated in Fig.~\ref{fig:fpf0} in which we 
compare the lattice results and the fit results for 
the vector and connected scalar cases.  

\subsubsection{Results including the disconnected contribution}
\label{sec:discscalar}

For the full scalar form factor we need to include the 
quark-line disconnected contribution from the lower 
diagram of Fig.~\ref{fig:2pt3ptdiagram}. 
We can then define flavour-singlet and flavour-octet 
scalar currents:
\begin{eqnarray}
\label{eq:sdef}
S_{\mathrm{singlet}} &=& 2 \overline{l}l + \overline{s}s  \\
S_{\mathrm{octet}} &=& 2 \overline{l}l - 2\overline{s}s . \nonumber
\end{eqnarray}
Scalar form factors for these two currents are then determined 
by combining the connected scalar form factor 
of Section~\ref{sec:connscalar} with disconnected 
contributions in appropriate combinations from 
quark loops made from $l$ quarks or $s$ 
quarks. 

\begin{figure}
\centering
\includegraphics{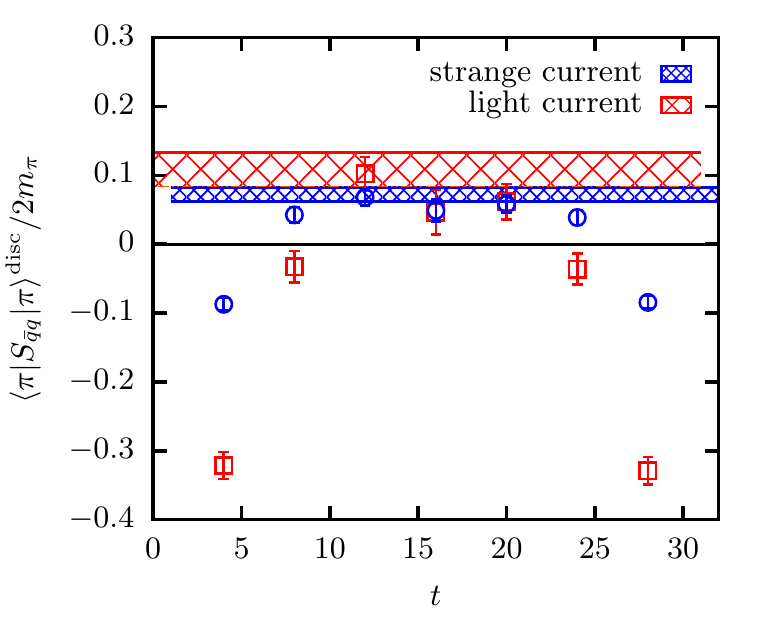}
\caption{The ratio of 3-point correlator to 2-point correlator
for the disconnected contribution for the $\overline{l}l$ 
(red circles) and $\bar{s}s$ currents (blue squares) to the scalar
form factor of the $\pi$ at $q^2=0$ on coarse lattices, set 2. 
The points are the lattice QCD results with statistical errors and the 
red and blue hashed bands show the ground-state fit result for
the $\overline{l}l$ and $\bar{s}s$ contributions, respectively. 
}
\label{fig:f0disc_ll}
\end{figure}

For the $q^2=0$ case it is particularly simple to calculate 
the disconnected contributions. Indeed, for the $\overline{s}s$ 
scalar current this is the $\pi$ meson 
equivalent of the `strangeness in the nucleon' 
calculation on which there has been a great deal work in 
lattice QCD (see, for example,~\cite{Freeman:2012ry}).
The disconnected contribution for current $\overline{q}q$ is 
\begin{equation}
\label{eq:disc}
\langle \pi | S_{\overline{q}q} | \pi \rangle^{\mathrm{disc}} = \langle \pi(p) | \overline{q} q | \pi(p) \rangle - \langle \pi(p) | \pi(p) \rangle \langle \overline{q}q \rangle .
\end{equation}
The first term is the ensemble average of a $\pi$ meson 2-point
function with source at time 0 and sink at time $T$ with a scalar 
current (condensate) insertion summed over the spatial points making 
up the timeslice at $t$. 
The second term in eq.~(\ref{eq:disc}) subtracts the 
product of the vacuum expectation values of the $\pi$ 
meson correlator and condensate. 

With HISQ quarks a convenient way to represent the $\overline{q}q$ 
condensate is as a  sum over a pseudoscalar meson correlator with valence 
quarks $q$~\cite{cond}. We use an identity~\cite{Kilcup:1986dg, cond} that relates the 
quark propagator for staggered quarks on a given gluon field configuration 
to a product of quark 
propagators summed over lattice sites: 
\begin{equation}
\label{eq:staggid}
\mathrm{Tr}M^{-1}_{00} = am_q \sum_n \mathrm{Tr}\left|M^{-1}_{0n}\right|^2 . 
\end{equation}
Here $0$ and $n$ are arbitrary lattice sites, $\mathrm{Tr}$ is a color trace
and $am_q$ is the quark mass 
in lattice units used in the propagator. 
The left-hand side of eq.~(\ref{eq:staggid}) is the negative of the quark loop from site 0 to 
site 0 needed for the disconnected piece of the scalar current and the 
the right-hand side is the Goldstone 
pseudoscalar meson correlator at zero spatial momentum for a quark-antiquark pair of mass 
$am_q$ multiplied by that mass. Since our Goldstone meson correlators here use a random-wall 
source a sum over a timeslice for lattice site $0$ for the quark loop is done implicitly. 

The quantities required to calculate the disconnected 
contribution for the $s\overline{s}$ current 
to the scalar form factor at $q^2=0$ are then simply 
$\pi$ meson correlators and those for the pseudoscalar 
$s\overline{s}$ meson known as the $\eta_s$. The $\eta_s$ 
does not correspond to a physical particle but its correlators 
are nevertheless usefully studied in lattice 
QCD~\cite{Davies:2009tsa} and so have been calculated 
previously~\cite{fkpi}. 
To make the 3-point function 
needed 
we take a $\pi$ meson correlation function with source at timeslice 
0 and sink at time-slice $T$ and a set of $\eta_s$ meson correlators 
with sources at time-slices denoted by $t$. 
For this calculation we use correlators made for the determination of 
$\pi$, $K$ and $\eta_s$ masses and decay constants in~\cite{fkpi}. 
These have zero spatial momentum and 16 time-sources, so that $t$ 
comes in steps of 4 timeslices on coarse set 2, which is 
the set that we will focus on. The $\overline{s}s$ current loop 
at time $t$ is obtained by summing over all end points for an 
$\eta_s$ correlator starting at
time-source $t$. 
For the disconnected contribution we multiply this by the $\pi$ 
correlator with time-source 0 and sink $T$, averaging over all 
time-sources on a configuration that give the same set of 
relative time separations. The 3-point function that 
yields $\langle \pi | S_{\overline{s}s} | \pi \rangle$ 
of eq.~(\ref{eq:disc}) at $q^2=0$ is thus given, 
averaging over gluon field configurations, by 
\begin{align}
\label{eq:disccorr}
C_{3pt}&(p_1,p_2,0,t,T) = \nonumber \\
&- \Big\langle C_{\pi}(p,0,T)am_s\sum_{t^{\prime}}C_{\eta_s}(p,t,t^{\prime}) \Big\rangle \nonumber\\
&+ \Big\langle C_{\pi}(p,0,T)\Big\rangle
\Big\langle am_s\sum_{t^{\prime}}C_{\eta_s}(p,t,t^{\prime}) \Big\rangle
\end{align}
where $p_1=p_2=p=0$.
For the scalar current made of light quarks an equivalent expression holds, 
using two $\pi$ meson correlators with offset time-sources. 

Figure~\ref{fig:f0disc_ll} shows results for a scalar current made 
of light quarks or strange quarks. The quantity plotted is the ratio
of the 3-point correlator 
generated from the equivalent of eq.~(\ref{eq:disccorr}) divided by 
the 2-point correlators for the $\pi$ meson at zero momentum
whose sources are at $T$ lattice spacings apart. 
We take $T = 32$ but have checked that results are very similar 
for other values of $T$, such as $T=28$. Because we are using 
point sources for our $\pi$ meson correlators we do not have a 
large plateau region. A longer plateau is obtained using smeared sources~\cite{priv}. 
However, by using a combined fit of the 3-point and 2-point 
correlators we can allow for systematic uncertainties from excited states 
and we obtain a good fit. The red and blue hashed bands show the fit results for 
the (unrenormalised) ground-state matrix element of the scalar current made of
light and strange quarks, respectively, divided by 
twice the $\pi$ meson mass. 

The results for the ground-state matrix elements for both the 
$l$ and $s$ scalar currents are given in Table~\ref{tab:disconnected}. 
At $q^2=0$ we can compare them to the connected contribution given 
in Table~\ref{tab:mep}. We see that the disconnected contributions are 
very much smaller, each around 1\% of the connected contribution. 
This is to be expected based on the Feynman-Hellmann 
theorem which would relate the disconnected contributions to the derivative 
of the $\pi$ meson mass with respect to the sea $s$ or $l$ quark 
mass~\cite{Freeman:2012ry}, in a similar way to that discussed in 
Section~\ref{sec:connscalar} for the connected contribution. 
Our results for the $s$ scalar current indicate reasonable agreement 
(within a factor of two) with the  
$\pi$ mass dependence on the $s$ sea quark mass (keeping all other 
parameters fixed) obtained at heavier-than-physical 
$\pi$ masses by the MILC collaboration~\cite{priv}. 

\begin{table}
\caption{
The $l$ and $s$ scalar current disconnected contributions to the scalar form factor 
on coarse set 2. 
}
\label{tab:disconnected}
\begin{ruledtabular}
\begin{tabular}{lll}
 $q^2/\mathrm{GeV}^2$ & $f_{0,\overline{l}l}^{\mathrm{disc}}(q^2)/Z_S$ & $f_{0,\overline{s}s}^{\mathrm{disc}}(q^2)/Z_S$ \\
\hline
0.0 & 0.0177(40)  & 0.0118(17)    \\
-0.0315 & -0.0152(67) &  0.0003(27)   \\
-0.0526 & -0.055(13) & -0.0078(72)    \\
\hline
\end{tabular}
\end{ruledtabular}
\end{table}

To obtain results for the disconnected contribution to the scalar 
form factor at non-zero values 
of $q^2$ we need to project onto non-zero lattice spatial 
momenta, $2\pi/L_s(n_x,n_y,n_z)$, at $T$ and $t$ in the correlators 
used in eq.~(\ref{eq:disccorr}). This extends eq.~(\ref{eq:staggid}) 
to the non-zero momentum case using translation invariance. 
The statistical errors grow as 
spatial momentum is introduced so we restrict ourselves to the 
smallest non-zero lattice momenta with $(n_x,n_y,n_z)$ equal 
to $(1,0,0)$ and $(1,1,0)$ and permutations thereof, including 
$-1$ as well as $+1$. We work only on coarse set 2. 

We obtain the ground-state matrix element for these contributions 
using a combined 3-point and 2-point fit as before. For this we 
need new 2-point $\pi$ meson correlators at these spatial 
momentum values and we obtain these on a subset of 600 configurations. 

\begin{figure}
\centering
\includegraphics{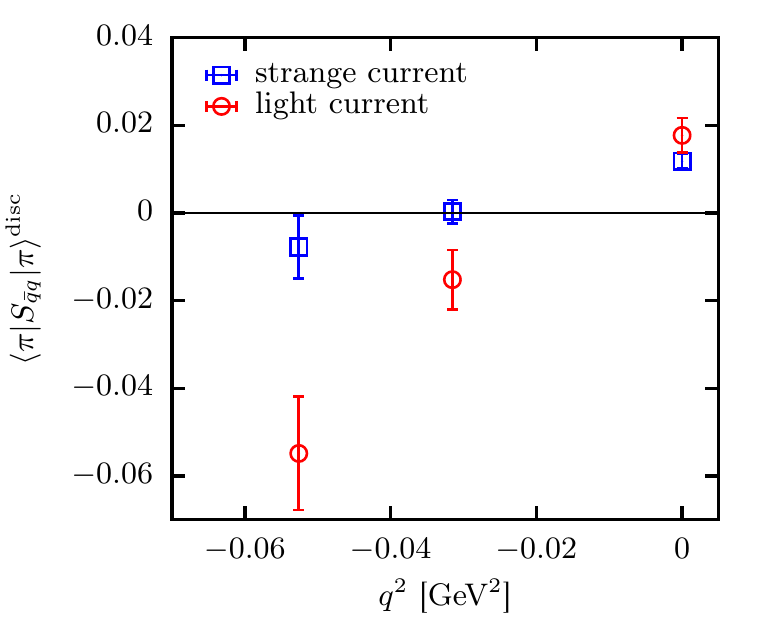}
\caption{The (unrenormalised) $s\overline{s}$ and $l\overline{l}$ disconnected 
contributions to the scalar form factor as a function of $q^2$ for 
coarse lattices, set 2. 
}
\label{fig:f0disc_q2}
\end{figure}

\begin{figure}
\centering
\includegraphics{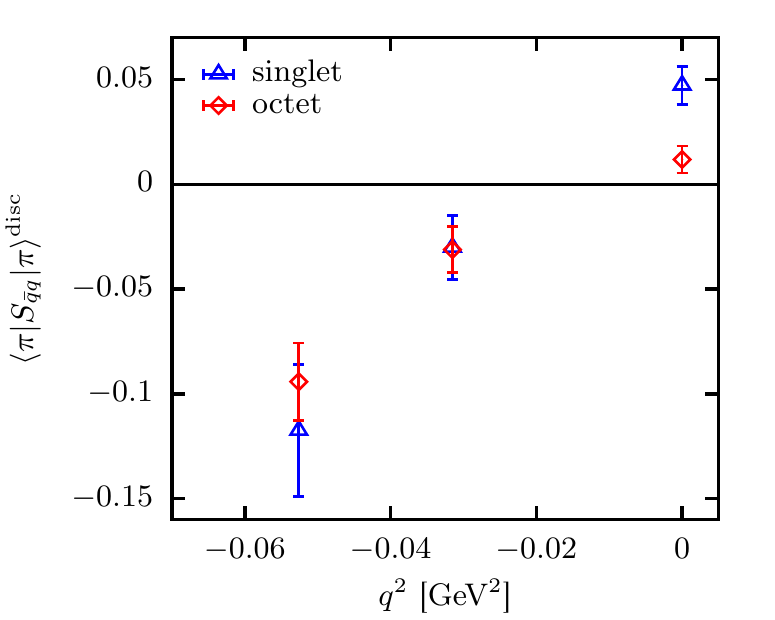}
\caption{The combination of $\overline{s}s$ and $\overline{l}l$ (unrenormalised) disconnected 
contributions to the scalar form factor that form either a flavour-singlet 
or a flavour-octet combination (see eq.~(\ref{eq:sdef})). These 
are plotted as a function of $q^2$ for 
coarse lattices, set 2. 
}
\label{fig:f0singlet_octet}
\end{figure}

Figure~\ref{fig:f0disc_q2} shows the disconnected contribution to 
the scalar form factor for $l$ and $s$ scalar currents, tabulated 
in Table~\ref{tab:disconnected}. It is clear from this plot that 
when disconnected contributions are included with a positive 
sign they will increase the slope of the form factor 
and therefore the mean square radius. Thus the flavour singlet scalar 
radius will be larger than the radius from the connected 
diagram only. This is less clear for the flavour octet case but   
the magnitude of the strange disconnected contribution is smaller than that of 
the light disconnected contribution so we might expect a net positive 
effect.  
Figure~\ref{fig:f0singlet_octet} collects the results into 
singlet and octet flavour combinations. Now it is clear that both singlet 
and octet radii will be larger than the radius from the connected diagram only.  

To obtain the mean square radius for the singlet and octet scalar form 
factors we must combine the connected and disconnected contributions. 
In doing this we must be careful to insert appropriate factors of 2 
for the $\overline{l}l$ pieces so 
that both the connected and disconnected contributions 
include $\overline{u}u+\overline{d}d$. 
Since we only have a calculation of the disconnected pieces on 
coarse set 2 we use a simple approach to determining the change 
in the mean square radius, using a linear approximation to the 
form factor over the small $q^2$ range (0 to -0.0315 $\mathrm{GeV}^2$) 
covered by the disconnected 
results. This has the advantage of making clear how the disconnected 
contributions affect the result. They appear both in the value of 
the total form factor at $q^2=0$ which is used for the normalisation 
and they contribute to the slope of the form factor in $q^2$. As discussed 
above, the effect on the form factor at $q^2=0$ is very small (1\%) and the 
largest effect comes from the contribution to the slope.  
We have, comparing the form factor at $q^2$ to that at 0,  
\begin{eqnarray}
\label{eq:rdisc}
\frac{|q^2|}{6}\langle r^2 \rangle &=& \frac{|q^2|}{6} \langle r^2 \rangle_{\mathrm{conn}}(1+f_{\mathrm{disc}}(0)/f_{\mathrm{conn}}(0))^{-1}  \\
&+& \frac{f_{\mathrm{disc}}(0)-f_{\mathrm{disc}}(q^2)}{f_{\mathrm{conn}}(0)}(1+f_{\mathrm{disc}}(0)/f_{\mathrm{conn}}(0))^{-1} . \nonumber
\end{eqnarray}
The second term makes a large contribution to the mean square radius because 
the change in the disconnected contribution to the form factor over the 
range in $q^2$ (depending 
on the combination of flavours) is of the 
same size as that of the connected contribution 
included in $\langle r^2\rangle_{\mathrm{conn}}$. 
We find, for example, that the change in 
mean square radius is 50(20)\% for the singlet 
combination. 

For the singlet and octet combinations we obtain: 
\begin{align}
\label{eq:r2scalardisc}
\langle r^2 \rangle_{S,\mathrm{singlet}}^{(\pi)} &= 0.506(38)(53) {\textrm{ fm}}^2 , \\  
\langle r^2 \rangle_{S,\mathrm{octet}}^{(\pi)} &= 0.431(38)(46) {\textrm{ fm}}^2 . \nonumber  
\end{align}
Here the first error is statistical and comes 
from adding in the disconnected contribution.  
The second error is systematic from electromagnetic/isospin 
and finite volume effects as discussed in Section~\ref{sec:connscalar} 
for the connected scalar radius. The full error budget for the singlet/octet 
radius is given in Table~\ref{tab:error}.

For comparison with earlier work on configurations that include only 
$u$ and $d$ quarks in the sea we can construct a radius that 
corresponds to the form factor for a $\overline{u}u+\overline{d}d$ 
scalar current. We find
\begin{equation}
\label{eq:r2ud}
\langle r^2 \rangle_{S,ud}^{(\pi)} = 0.481(37)(50) {\textrm{ fm}}^2 .   
\end{equation}

As eq.~(\ref{eq:rdisc}) makes clear, the results for the different 
scalar radii are correlated. The differences between them 
are significant since a lot of the uncertainty cancels. 
For example
\begin{equation}
\label{eq:r2diff}
\langle r^2 \rangle_{S,\mathrm{singlet}}^{(\pi)} - \langle r^2 \rangle_{S,\mathrm{octet}}^{(\pi)} = 0.075(20) {\textrm{ fm}}^2 .  
\end{equation}
We find the ordering:
\begin{equation}
\label{eq:r2order}
\langle r^2 \rangle_{S,\mathrm{singlet}}^{(\pi)} > \langle r^2\rangle_{S,ud}^{(\pi)} > \langle r^2 \rangle_{S,\mathrm{octet}}^{(\pi)} > \langle r^2 \rangle_{S,\mathrm{conn}}^{(\pi)} .  
\end{equation}

%%%%%%%%%%%%%%%%%%%%%

\section{Discussion} 
\label{sec:discussion}

\begin{figure}
\centering
\includegraphics{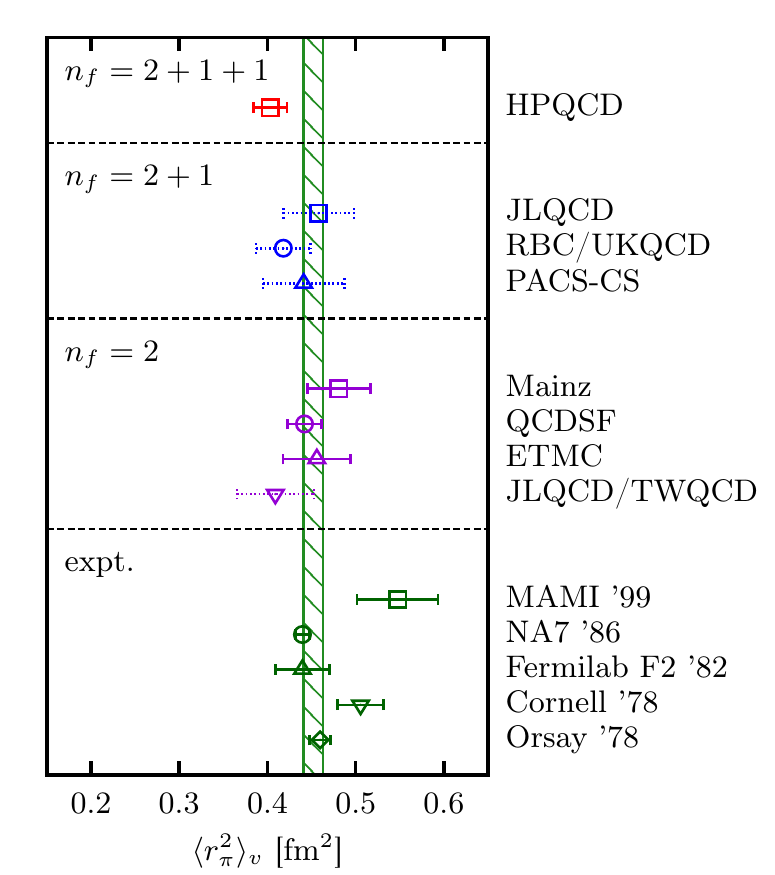}
\caption{A summary of lattice QCD results for the mean square electric 
charge radius of the $\pi$ meson arranged by the number of quark 
flavours included in the sea. 
The top result is from this paper; those including $u$, $d$, and 
$s$ quarks in the sea ($n_f=2+1$) are from~\cite{rbcukqcd, Nguyen:2011ek, Aoki:2015pba} 
and those including only $u$ and $d$ quarks in the sea ($n_f=2$) 
are from~\cite{Brandt:2013dua, qcdsf, etmpiff, Aoki:2009qn}. 
Results that include only one value of the lattice spacing have dotted 
error bars. 
Experimental results are from~\cite{Liesenfeld:1999mv, amendolia, Quenzer:1978qt, dally:1982zk, Bebek:1977pe}. 
The hashed vertical line gives the average from 
the Particle Data Group~\cite{PDG2014}. 
}
\label{fig:r2v_sum}
\end{figure}

\begin{figure}
\centering
\includegraphics{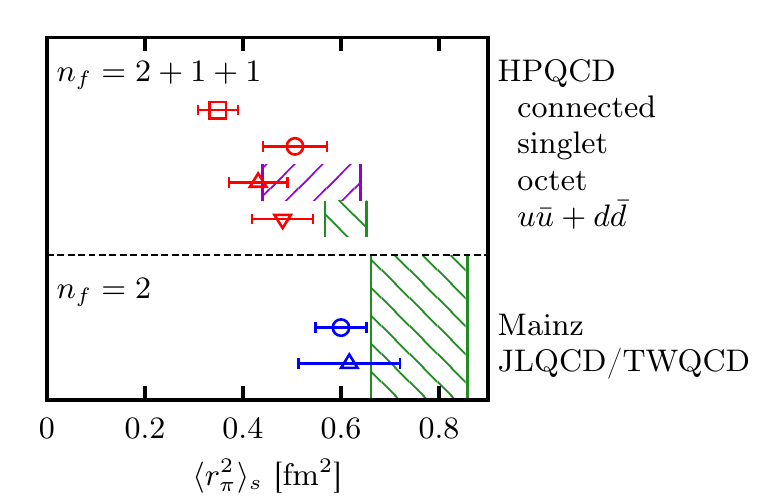}
\caption{A summary of lattice QCD results for the mean square scalar
radius of the $\pi$ meson arranged by the number of quark 
flavours included in the sea. The HPQCD Collaboration's  results are from
this paper: ``connected'' shows the mean square radius from the quark-line connected
calculation only (eq.~(\ref{eq:rsres})); 
``singlet'' and ``octet'' are full calculations including
quark-line disconnected diagrams arranged in flavour-singlet or flavour-octet
currents (see eqs.~(\ref{eq:sdef}) and~(\ref{eq:r2scalardisc})) 
and $\overline{u}u+\overline{d}d$ 
includes only $u/d$ quarks in the scalar current (eq.~(\ref{eq:r2ud})).
The results including only $u$ and $d$ quarks in the sea ($n_f=2$) 
are from~\cite{Gulpers:2015bba,Aoki:2009qn}. The hashed green vertical bands 
give the result expected from chiral perturbation theory for $F_{\pi}/F$ 
for $n_f=2$ and $n_f=2+1$ (for comparison with our $n_f=2+1+1$ results) 
from eq.~(\ref{eq:fpif}). 
The phenomenological result from $\pi-\pi$ scattering~\cite{gasserl} is 
very similar to the $n_f=2+1$ green band.
The hashed purple vertical band gives the chiral perturbation 
theory expectation for the scalar octet mean square 
radius (eq.~(\ref{eq:r2seval}))~\cite{Gasser:1984ux}. 
}
\label{fig:r2s_sum}
\end{figure}

Figure~\ref{fig:r2v_sum} compares the result obtained in this paper for the
mean square of the pion electric charge radius to other lattice QCD calculations by
RBC/UKQCD~\cite{rbcukqcd}, PACS-CS~\cite{Nguyen:2011ek}, the Mainz group~\cite{Brandt:2013dua},
QCDSF~\cite{qcdsf}, ETMC~\cite{etmpiff} and JLQCD/TWQCD~\cite{Aoki:2009qn, Aoki:2015pba}, and
to experimental results~\cite{Liesenfeld:1999mv, amendolia, Quenzer:1978qt, dally:1982zk, Bebek:1977pe}. It should be noted that several of these calculations include results at 
only one value of the lattice spacing and error budgets are not complete in all cases. 
A recent calculation by B. Owen \textit{et. al.}~\cite{Owen:2015gva}
used one lattice spacing and five different pion masses down to
$156$~MeV but no chiral or continuum extrapolation is given so the results
are not included in the figure.

The calculation presented in this paper is the first one that has been done at
the physical pion mass --- other lattice QCD calculations have used heavier
than physical pions. However, as Figure~\ref{fig:r2v_sum} shows, all lattice
QCD results agree well after extrapolation to zero lattice spacing and physical
pion mass. We see no difference between the lattice calculations using
different sea quark content ($u$ and $d$ only, $u$, $d$ and $s$, or $u$, $d$,
$s$ and $c$ quarks in the sea) at this level of accuracy.
In Figure~\ref{fig:fvsexp} we compare the shape of the electromagnetic form factor
from our calculation to the result by NA7~\cite{amendolia}, which is the most accurate
one of the experimental results. The agreement is good, which shows also here
when we compare our result for the mean square radius to the NA7 results.
Our value is 2$\sigma$ below the average of the experimental results~\cite{PDG2014}. 

In the case of the scalar radius, comparison with other lattice QCD results must 
be done with care.  
Two lattice QCD calculations have been done including $u$ and $d$
quarks in the sea ($n_f=2$). These are by the Mainz group~\cite{Gulpers:2015bba}, recently 
updating~\cite{Gulpers},
and the JLQCD/TWQCD collaborations~\cite{Aoki:2009qn}. Both of these calculations
include the quark-line disconnected diagrams but only for a $\overline{u}u+\overline{d}d$ 
scalar current (consistent within an $n_f=2$ framework). 
Here we include $u$, $d$, $s$ and $c$ quarks in the sea and a scalar current that 
includes also $\overline{s}s$ contributions in two different overall flavour 
combinations. We neglect $\overline{c}c$ contributions 
to the current since we expect those contributions to be suppressed by 
powers of the $c$ quark mass. 

The pion scalar form factor is not
directly accessible to experiment as there is no suitable low-energy probe. However,
the scalar radius can be determined from the cross section for $\pi-\pi$ scattering
and from the pion decay constant by using chiral perturbation theory~\cite{gasserl, Gasser:1984ux}. 
In SU(2) chiral perturbation theory the ratio of the physical pion decay constant to that 
in the chiral limit
can be related to the pion scalar radius for the $\overline{u}u+\overline{d}d$ scalar current 
by~\cite{gasserl}
\begin{equation}
\frac{F_{\pi}}{F}=1+\frac{m^2_{\pi}}{6}\langle r^2\rangle^{(\pi)}_{S,ud}+\frac{13}{192\pi^2}\frac{m^2_{\pi}}{F^2_{\pi}}
+\mathcal{O}(m^4_{\pi})
\end{equation}
where $F_{\pi}=f_{\pi}/\sqrt{2}$=92 MeV and we take $m_{\pi}$ = 135 MeV. 
$F_{\pi}/F$ values from SU(2) chiral 
perturbation theory analyses of lattice QCD calculations are collected in~\cite{flag} 
and give averages:
\begin{eqnarray}
\label{eq:fpif}
\frac{F_{\pi}}{F} &=& 1.0744(67), \, \langle r^2\rangle^{(\pi)}_{S,ud} = 0.76(9)(4) \mathrm{fm}^2, \, n_f = 2  \\
 &=& 1.0624(21), \, \langle r^2\rangle^{(\pi)}_{S,ud} = 0.61(3)(3) \mathrm{fm}^2, \, n_f = 2+1 \nonumber 
\end{eqnarray} 
where we have included a second uncertainty of 5\% for higher order corrections 
to the chiral perturbation theory formula. The results for $n_f=2$ and $n_f=2+1$ 
analyses are compatible within 2$\sigma$ but do not have to agree. A phenomenological 
estimate for $\langle r^2\rangle^{(\pi)}_{S,ud}$ based on $\pi-\pi$ scattering
gives 0.61(4) $\mathrm{fm}^2$~\cite{Colangelo:2001df}, which agrees well with the $n_f=2+1$ 
result above. For our $n_f=2+1+1$ calculations we compare our value 
for $\langle r^2 \rangle^{(\pi)}_{S,ud}$ to the $n_f=2+1$ results, 
following the discussion of the compatibility of $n_f=2+1$ and $n_f=2+1+1$ chiral 
analyses in~\cite{flag}.  

When $\overline{s}s$ components are included in the current we can form flavour octet 
and singlet combinations. The flavour octet combination is interesting because 
it can be estimated from $f_K/f_{\pi}$ since no new 
low-energy constants appear~\cite{Gasser:1984ux}. The mean square radius is given by 
\begin{equation}
\label{eq:r2octchi}
\langle r^2\rangle^{(\pi)}_{S,\mathrm{octet}} = \frac{6}{m_K^2-m_{\pi}^2}\left( \frac{F_K}{F_{\pi}}-1 \right) + \delta_3 
\end{equation} 
with
\begin{eqnarray}
\label{eq:delta3}
\delta_3 &=& \frac{1}{64\pi^2F_{\pi}^2}\frac{1}{m_K^2-m_{\pi}^2}\left\{ 6(2m_K^2-m_{\pi}^2)\ln \frac{m_K^2}{m_{\pi}^2} \right. \nonumber \\
&+& \left. 9m_{\eta}^2\ln \frac{m_{\eta}^2}{m_{\pi}^2} -2(m_K^2-m_{\pi}^2)\left[ 10+\frac{m_{\pi}^2}{3m_{\eta}^2}\right] \right\}. 
\end{eqnarray}
Using $m_{\pi} = 135 \mathrm{MeV}$, $m_K = 496 \mathrm{MeV}$, $m_{\eta} = 548 \mathrm{MeV}$ 
and $F_{K}/F_{\pi} = 1.1916$~\cite{fkpi} gives 
\begin{equation}
\label{eq:r2seval}
\langle r^2\rangle^{(\pi)}_{S,\mathrm{octet}} = 0.54(10) \mathrm{fm}^2
\end{equation} 
where we take the estimate from~\cite{Gasser:1984ux} of the uncertainty from higher order 
terms. Note that we expect this mean 
square radius to be smaller than 
$\langle r^2 \rangle^{(\pi)}_{S,ud}$ because it involves the subtraction 
of twice the strange current quark-line disconnected contribution, and our 
results show this to have a positive impact on the mean square radius.  

The difference between singlet and octet mean square radii comes 
from the $q^2$ dependence of the 
matrix element of the $\overline{s}s$ piece of the 
scalar current. It is 
denoted $\langle \delta r^2 \rangle_S$ in~\cite{Gasser:1984ux} 
and estimated in chiral perturbation theory as:
\begin{eqnarray}
&& \langle \delta r^2 \rangle_S =  \\
&& \frac{6}{F_{\pi}^2} \left( 12L_4^r - \frac{3}{64\pi^2}\left[ \ln \frac{m_K^2}{\mu^2}+1\right] +\frac{m_{\pi}^2}{288\pi^2 m_{\eta}^2} \right) . \nonumber
\end{eqnarray}
Using $L_4^r$ renormalised at $m_{\eta}$ from~\cite{fkpi} gives 
$\langle \delta r^2 \rangle_S = 0.015 \pm 0.1 \,\mathrm{fm}^2$ which, given 
its uncertainty, agrees with our result in eq.~(\ref{eq:r2diff}). 

\begin{figure}
\centering
\includegraphics[width=\hsize]{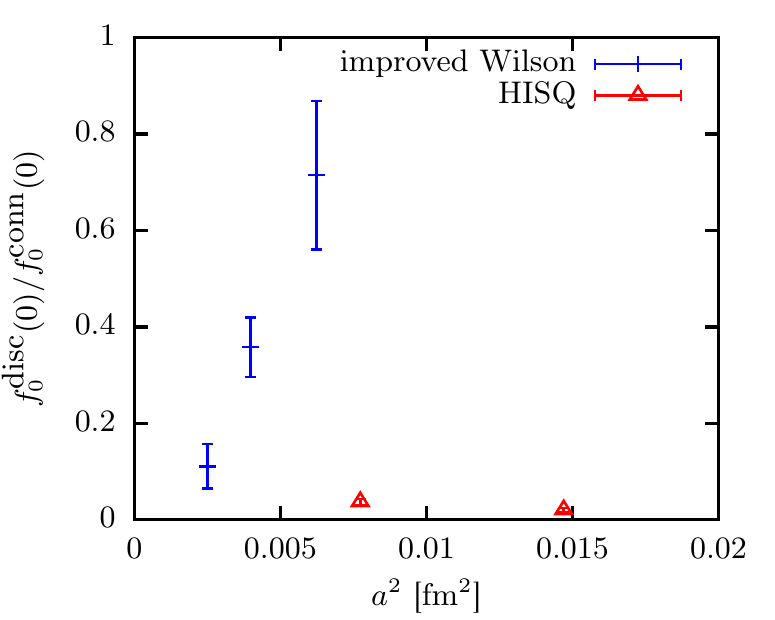}
\caption{A comparison of the ratio of quark-line disconnected to 
connected contributions for the $\overline{u}u+\overline{d}d$ current 
to the pion scalar form factor
at $q^2=0$ plotted against the square of the lattice spacing. 
Results denoted by the blue plus symbol for an improved Wilson action 
are taken from~\cite{Gulpers:2015bba}. 
The red triangles are from the results presented here for the HISQ 
action. 
}
\label{fig:disc-compare}
\end{figure}

Figure~\ref{fig:r2s_sum} compares our results for our various 
pion scalar mean square radii 
to those obtained on $n_f=2$ gluon 
configurations and to the expected values from chiral perturbation theory 
given above. 
There is reasonable agreement (within $2\sigma$) between all the lattice QCD results for 
$\langle r^2\rangle^{(\pi)}_{S,ud}$ and with the values expected from $F_{\pi}/F$.
Our result for the flavour octet radius is also in good agreement with 
the value in eq.~(\ref{eq:r2seval}). 
To illustrate how
important the contributions from the disconnected diagrams are to the various scalar radii
we also show the result from our calculation
of the connected diagram only (eq~(\ref{eq:rsres})). 
As discussed in Section~\ref{sec:discscalar} 
the contributions from the disconnected diagrams to the form
factor are small but the change in the slope and therefore in the radius 
is substantial. This feature of the scalar radius is also 
discussed in~\cite{Gulpers, Gulpers:2015bba}.

Our results are in both qualitative agreement and reasonable 
quantitative agreement with the picture expected from chiral perturbation 
theory~\cite{Juttner:2011ur}. There the disconnected contribution 
to the form factor is predicted to be very small at $q^2=0$, becoming 
negative at negative $q^2$ values so that the contribution to the 
radius is substantial (approximately equal to that of the connected 
contribution) and positive. We find the contribution to amount to 
an approximately 50\% increase in the radius, rather than doubling, 
but with substantial uncertainty.  

Although different lattice QCD formalisms 
will have a different normalisation for the scalar current, 
the ratio of disconnected to connected contributions to the form 
factor at a given value of $q^2$ should agree in the chiral and 
continuum limits 
since renormalisation factors will then cancel. 
Our results obtained here with the HISQ action seem to agree well with 
those from the overlap action given at one value of the 
lattice spacing in~\cite{Aoki:2009qn}, judging this from their Figure 9. 
They do not agree well with those from an improved Wilson action given 
at three values of the lattice spacing in~\cite{Gulpers:2015bba}. 
They have a very substantial disconnected contribution that also 
shows, as a ratio of the connected contribution, a very strong 
lattice spacing dependence. 
They work at heavier values of $m_{\pi}$ than 
we do but see little dependence on $m_{\pi}$ in this ratio 
(apart from one point that they suggest to treat as an outlier). 
In Figure~\ref{fig:disc-compare} we show a comparison of their 
disconnected/connected ratio at $q^2=0$ to ours. For their points 
we have taken numbers from their Figure 2, using the values closest 
to $m_{\pi}$ = 250 MeV (so any variation with $m_{\pi}$ is not 
included in our plot, and it should be taken as an approximate 
representation of their results). For our results we give values from 
our coarse set 2 and from fine set 3.  Although our results also 
show some lattice spacing dependence, it is hard to avoid the 
conclusion that the improved Wilson results are dominated by a 
lattice artefact. It is not clear whether the values from
HISQ/domain wall and improved Wilson actions will 
agree in the continuum (and chiral) limits and this does need to be resolved.  

\section{Conclusions} 
\label{sec:conclusions}

We have given the first lattice QCD results for the vector and scalar 
form factors of the $\pi$ meson including $u/d$ quarks with their physical 
masses. 

Our results for the vector form factor as a function of $q^2$ 
agree well (needing no extrapolation) with the experimental 
values from $\pi-e$ scattering (see Figure~\ref{fig:fvsexp} and 
eq.~(\ref{eq:rvres})). This confirms the encouraging picture 
seen by earlier lattice QCD calculations (albeit with heavier $u/d$ quark 
masses and a less realistic QCD vacuum) that lattice QCD does indeed 
reproduce the QCD effects that result in a finite electric charge radius 
for the $\pi$ meson. It would clearly be possible to extend our 
results to higher values of $q^2$ with the aim of eventually 
matching on to expectations from QCD perturbation theory 
(see, for example,~\cite{lhp}). This would require finer lattices 
to avoid large discretisation errors from the discretisation of 
momentum and, for numerical efficiency, should probably be done 
with heavier-than-physical $\pi$ mesons (even $\eta_s$ mesons). 

The $\pi$ scalar form factor is of less immediate phenomenological 
interest but can be stringent test of low-energy expectations from 
QCD and from chiral perturbation theory (where it can be related 
to decay constants and $\pi-\pi$ scattering). 
Here we have given the first results to include $u$, $d$ and $s$ quarks 
in the sea and in the scalar current. This allows us to define 
a number of different radii for different flavour combinations. 
Calculation of the scalar form factor must include the effect 
of quark-line disconnected diagrams and we agree with earlier results 
that these have a substantial impact on the determination of 
the radii. An increase in the radius occurs where $\overline{q}q$ 
has a positive coefficient in the combination that appears in the 
scalar current and we find the magnitude of the contribution of 
$\overline{l}l$ to be larger than that of $\overline{s}s$. 
We therefore have an ordering in value of the different radii 
that we give in eq.~(\ref{eq:r2order}). 
Our values for the quark-line disconnected contribution 
to the form factor are, however, very small at $q^2=0$ in 
agreement with expectations from chiral perturbation 
theory and with earlier results using the overlap formalism~\cite{Aoki:2009qn}. 

Our value for the radius obtained using a $u/d$ scalar current 
(eq.~(\ref{eq:r2ud})) agrees 
within $2\sigma$ with expectations from chiral perturbation 
theory and earlier values from calculations including $u$ and 
$d$ quarks in the sea. 
We give the first results for the radius from the flavour octet 
and flavour singlet
scalar currents (eq.~(\ref{eq:r2scalardisc})). 
The flavour octet mean square radius agrees well with expectations from chiral 
perturbation theory where it can be related to $f_K/f_{\pi}$ and 
combinations of meson masses. 

Our largest source of uncertainty in 
the scalar case is from finite-volume corrections since we are working 
with physically light $\pi$ mesons, even if on lattices 5.8 fm across. 
To improve uncertainties on the scalar radius we would need to use 
larger volumes at the physical point, or include a dedicated study of 
finite-volume effects away from the physical point, and ensembles of gluon
field configurations exist on which 
this could be done. 
Improved statistics on the quark-line disconnected 
contributions are also necessary to reduce the statistical uncertainty 
in their impact on the scalar mean square radius. 
The new techniques we have introduced here for handling the disconnected 
contributions in the pion scalar form factor make these improvements feasible in
future results. 

{\it Acknowledgements.} We are grateful to MILC for the use of their 
gauge configurations and code and to G. Donald, W. Freeman, V. G\"{u}lpers, A. Lytle, D. Toussaint  
and G. von Hippel for useful discussions. 
This work was funded by STFC, NSF, the Royal Society and the Wolfson 
Foundation. 
We used the Darwin Supercomputer 
of the University of Cambridge High Performance Computing Service 
as part of STFC's DiRAC facility jointly funded by STFC, BIS and 
the Universities of Cambridge and Glasgow. 
We are grateful to the Darwin support staff for assistance.

\bibliography{piff}

\end{document}